\documentclass{birkjour}
 \theoremstyle{definition}
 
 \theoremstyle{remark}

 \numberwithin{equation}{section}

\usepackage{graphicx}
\usepackage{subfig}
\usepackage{amssymb}
\usepackage{amsmath}
\usepackage{bm}
\usepackage{slashbox}
\usepackage{color}
\usepackage{slashed}
\newcommand{\bme}{\mathbf{e}}
\newcommand{\bea}{\begin{eqnarray}}
\newcommand{\eea}{\end{eqnarray}}
\newcommand{\bege}{\begin{equation}}
\newcommand{\enge}{\end{equation}}
\newcommand{\beq}{\begin{eqnarray}}
\newcommand{\benu}{\begin{enumerate}}
\newcommand{\enu}{\end{enumerate}}

\newcommand{\eeq}{\end{eqnarray}}

\newcommand{\mt}{\mathcal}

\newcommand{\CC}{\mathbb{C}}

\begin{document}

%-------------------------------------------------------------------------
% editorial commands: to be inserted by the editorial office
%
%\firstpage{1} \volume{228} \Copyrightyear{2004} \DOI{003-0001}
%\seriesextraline{This is the Concrete Title of this Book\br H.E. R and S.T.C. W, Eds.}
%\firstpage{1}

\title[Exotic Spinorial Structure]
 {Exotic Spinorial Structure and Black Holes in General Relativity}

%----------Author 1
\author[D. Beghetto]{Beghetto, D.}

\address{%
Departamento de F\'isica e Qu\'imica, Universidade Estadual Paulista (UNESP)\\
Av. Ariberto Pereira da Cunha, 333\\
CEP 12516-410\\
Guaratinguet\'a, SP\\
Brazil}

\email{dbeghetto@feg.unesp.br}

%\thanks{This work was completed with the support of our
%----------Author 2
\author{Cavalcanti, R. T.}
\address{%
Departamento de F\'isica e Qu\'imica, Universidade Estadual Paulista (UNESP)\\
Av. Ariberto Pereira da Cunha, 333\\
CEP 12516-410\\
Guaratinguet\'a, SP\\
Brazil}
\email{rogerio.txc@feg.unesp.br}

%----------Author 3
\author{Hoff da Silva, J. M.}
\address{%
Departamento de F\'isica e Qu\'imica, Universidade Estadual Paulista (UNESP)\\
Av. Ariberto Pereira da Cunha, 333\\
CEP 12516-410\\
Guaratinguet\'a, SP\\
Brazil}
\email{hoff@feg.unesp.br}


\date{July 1, 2018}

\begin{abstract}
We explore different (and complementary) views of spinors and their exotic counterparts, linking the very existence of the later to the presence of black holes. Moreover, we investigate the effects of the exotic term in the Hawking radiation emission rate, as well as its extremes, for asymptotically flat black holes solutions of general relativity. We show that, under certain circumstances, the emission rate extreme condition fixates an equation from which the exotic term could be inferred.
\end{abstract}

%%% ----------------------------------------------------------------------
\maketitle
%%% ----------------------------------------------------------------------
%\tableofcontents
\section{Introduction}

It is almost impossible to give the right emphasis on the importance of spinor fields in the construction of our understanding of the physical phenomena. Spinor fields constitute an irremediable essential tool without which high energy physics cannot be described. By the same token, it is equally hard to envisage the usefulness of the comprehensive mathematical construction behind the very concept of spinors. In fact, the general algebraic structure supporting the spinor concept has abundant applications in physics, from condensed matter to cosmology, going through quantum field theory. In spite of that, the construction of a solid bridge between a well established concept at the algebraic level and its counterpart in physics is highly nontrivial. In other words, the precise use of spinors in physics is hardly reached when the departure point is the algebraic definition of the spinorial quantity in question.    

Obviously, the point raised in the previous paragraph is not relevant within a more pragmatic point of view. In some cases, it is indeed irrelevant whether the spinor is understood as a mathematical object belonging to a section of a specific fiber bundle or as a four component ``column'', as far as it helps to give an observable agreeing with experience in a good level. However, a pragmatic approach, while quite satisfactory and acceptable in some cases, is not always free of the brevity narrowness. In several cases the formal emphasis lead to possibilities other than the highlighted by means of the usual thought. 

In the formal mathematical treatment concerning spinors, two seminal approaches are of particular interest in physics: the concept of spinors as ``pre-geometric'' quantities, in a manner of speaking, pointed out by Cartan \cite{ECARTAN}, in which the spacetime points themselves can be understood as generated by spinors components (a program which has been generalized eminently by Penrose \cite{Pe84}) and, on the other hand, the profound algebraic view of spin structures in a given manifold, codified in the \v{C}ech cohomology class \cite{naka}. 

Apart from its mathematical rigidity, both approaches may be connected in a tentative intuitive fashion providing an interesting picture to explore exotic spinors in a physical ground. We shall explore this connection in this paper, further developing this point of view and applying the resulting construction in the investigation of a physical system which we believe may serve as an interesting laboratory to explore questions about nontrivial topology, its changes, and impact in high energy physics. The idea is to explore at a physical level a consequence of linking the (time) variation of a specific additional term appearing in the connection (hence, the dynamics) of exotic fermions, and the emission of radiation by black hole through the Hawking process. Even further, by understanding the very existence of black holes as inductors of nontrivial topologies, we investigate how the extremes of emission rate of exotic spinors are influenced by nearly adiabatic changes in the black hole area.

The present paper is organized as follows: Section II is devoted to the study of exotic spinors, not only reviewing its formal aspects but also extending an intuitive approach to this subject. In Section III we investigate a physical output of a time-variable topology performed by a slowly varying black hole radius, analyzing the Hawking radiation emission rate in this case.  In order to clarify the different physical effects of gauge and exoticness, in the Appendix \ref{append} we show that the exotic term induces modifications on the dispersion relation incompatible with the ones due to the gauge field. In the final Section we conclude and discuss the results.

\section{Exotic spinors}

In the case of a nontrivial topology for the base manifold, there is not one but many spinorial structures different from the usual. Spinors belonging to these additional structures have, as a net effect of the nontrivial topology, a slightly different dynamics, as the derivative operator encodes a new term coming from such non triviality. We shall explore two approaches to this key correction. The first one is basically a short review on the formal aspects of exotic spinors\footnote{See Ref. \cite{exo} for a more complete account.}. The idea is to pinpoint the strictly necessary formalism to evince the existence and crucial effects related to the exoticity.  Subsequently, we also revisit an intuitive approach to exotic spinors, taking advantage of the Cartan's systematic approach to spinorial quantities, which will be further explored here. %%aqui

\subsection{Essential Formal aspects, a very brief review}

Not only one spin structure arise from different and inequivalent patching of local coverings for a given manifold $M$ \cite{Hawking:1977ab,Geroch:1968zm,Geroch:1970uv} . These possible nontrivial spin structures are labeled by elements of the well known group of homomorphisms of $\pi_1(M)$ into $\mathbb{Z}_2$, i. e., the first cohomology group $H^1(M,\mathbb{Z}_2)$ \cite{trinta,trintaseis,Asselmeyer}. Although the mathematical perspective is intricate, some basilar aspects may be pinpointed. For instance, when dealing with a simply connected base manifold, the fundamental group in indeed trivial and, hence, there is only one spin structure. In multiple connected manifolds, however, the situation is quite peculiar.  

Let us start with an appropriate definition of a spin structure on $M$. It is composed by a principal fiber bundle, say $\pi_s$, from the orthonormal coframe bundle $P_{Spin_{1,3}}(M)$ to $M$ along with the two-fold covering $s: P_{Spin_{1,3}}\rightarrow P_{SO_{1,3}}$ in such a way that $\pi_s=\pi\circ s$ for $\pi: P_{SO_{1,3}}\rightarrow M$. The assertion of the previous paragraph may be read in the following terms: if $H^1(M,\mathbb{Z}_2)\neq 0$ (notation for non-triviality), then different spin structures $(\tilde{P}_{Spin_{1,3}},\tilde{s})$ are in order. Noticed that a non-triviality on $M$ leads to an explicit dependence on different base manifold coverings for the definition of $\pi$, $\pi_s$. This dependence gives rise, in a manner of speaking, to its tilde counterparts. Now, the orthonormal coframe bundle act as a substrate for the covariant spinor bundle of $P_{Spin_{1,3}}(M)\times_\rho \mathbb{C}^4$ of which a classical spinor $\psi$ is a section. Therefore, it is indeed expected the existence of $\tilde{\psi}\in \sec \tilde{P}_{Spin_{1,3}}(M)\times_\rho \mathbb{C}^4$. In the notation here adopted, $\rho$ stands for the Weyl representation space $(1/2,0)\oplus(0,1/2)$ as a whole (or, eventually, to a single part of it).  

The equivalence between two given spin structures $(P_{Spin_{1,3}},s)$ and $(\tilde{P}_{Spin_{1,3}},\tilde{s})$ is performed by means of a mapping $q:P_{Spin_{1,3}}\rightarrow \tilde{P}_{Spin_{1,3}}$, such that $q=s\circ \tilde{s}$. Take now a point $x\in M$ and let $U_i$ and $U_j$ be two open sets on $M$ whose intersection encompass $x$. From $U_i\cap U_j\subset M$ it is possible to define two mappings $m_{ij}$ and $\tilde{m}_{ij}$ such that 
\begin{eqnarray}
U_i\cap U_j\subset M \underbrace{\longrightarrow}_{m_{ij}} Spin_{1,3} \hspace{2cm} U_i\cap U_j\subset M \underbrace{\longrightarrow}_{\tilde{m}_{ij}} \tilde{Spin}_{1,3}, \label{1}
\end{eqnarray} with a common transition, say $\sigma$, to $SO_{1,3}$. An additional mapping can be defined \cite{map} such that $m_{ij}(x)=\tilde{m}_{ij}(x)c_{ij}$ with the requirement that $c_{ij}: U_i\cap U_j\rightarrow \mathbb{Z}_2\hookrightarrow Spin_{1,3}$. 

In order to define a meaningful covariant derivative, it is generally assumed the existence of unimodular functions $\xi_i:U_i \subset M \rightarrow \mathbb{C}$ such that $\xi_i(x)\in U(1)$ \cite{trinta,trintaseis}. In fact, in the absence of torsion, the existence of such functions is guaranteed \cite{trinta,novo}. As we have mentioned, a spinor field $\psi$ is an element of $\sec P_{Spin_{1,3}}(M)\times_\rho \mathbb{C}^4$. However, different patching of local coverings resulting from nontrivial topology are labeled by different elements of $H^1(M, \mathbb{Z}_2)$. To each element of $H^1(M, \mathbb{Z}_2)$ it is associated a connection $\nabla$, giving rise to a one-to-one correspondence between inequivalent spin structures (and consequently inequivalent spinor fields) and elements of the first cohomology group. Having said that, it is a matter of working through the definition of a bundle mapping 
\begin{eqnarray}
f:\tilde{P}_{Spin_{1,3}}\times_\rho \mathbb{C}^4 &\rightarrow& P_{Spin_{1,3}}\times_\rho \mathbb{C}^4\nonumber\\
\tilde{\psi}&\mapsto &q(\tilde{\psi})=\psi,\label{uia}
\end{eqnarray} in such a way that 
\begin{eqnarray}
\tilde{\nabla}_X f(\tilde{\psi})=f(\nabla_X\tilde{\psi})+\frac{1}{2}(X, \xi^{-1}d\xi)f(\tilde{\psi}),\label{eia}
\end{eqnarray} for all $\psi \in  P_{Spin_{1,3}}\times_\rho \mathbb{C}^4$ for all vectorial field $X$ consistently defined over $M$. In Eq. (\ref{eia}) the brackets in the second term stands for an specific contraction not important to our general argumentation. The relevant point is that the unimodular field engenders an additional term to the covariant derivative, coming genuinely from the nontrivial topology. 

\subsection{Intuitive aspects} 

We intend here to give a complete account of an intuitive argument firstly presented in Ref.\footnote{A special collection, entitled ``Open Questions in Black Hole Physics'', Ed. Gonzalo J. Olmo.} \cite{uni}, along with a proper generalization. The underlying idea is to argue on the derivative term correction coming from the nontrivial topology. We would like to revisit and extend this intuitive point of view in this very section. Starting from the Cartan's spinorial approach we motivate the insertion of the second term in (\ref{eia}). Also, at the end of this section, bearing in mind that a black hole itself may be the inductor of a nontrivial topology, we make a link with the idea we are going to explore in the next section. 

A spacetime vector ${\bf v}\in\mathbb{R}^{1,3}$ can be expressed by ${\bf v} = x^0 \bme_0 + x^1 \bme_1 + x^2 \bme_2 + x^3 \bme_3,$ where ($x^0, x^1, x^2, x^3$) denote components of ${\bf v}$ with respect to an orthonormal basis  \{$\bme_0, \bme_1, \bme_2, \bme_3$\}. Null vectors \index{null vector}are isotropic vectors, and  satisfy $(x^0)^2 - (x^1)^2 - (x^2)^2 - (x^3)^2 = 0$. They  present null  directions in $\mathbb{R}^{1,3}$ with respect to the origin $\mathcal{O}$ of an arbitrary frame in $\mathbb{R}^{1,3}$. The space of null directions \index{null direction}that are future [past] pointed\footnote{Such space is nothing but a Riemann sphere.} are denoted by $\mathcal{S}^+$   [$\mathcal{S}^-$], and represented by the  intersections $\mathbb{E}^+$ [$\mathbb{E}^-$] of the future [past] light cones with the hyperplanes  $x^0 = 1$ [$x^0 = -1$]. The space $\mathbb{E}^\pm$ is a sphere with equation $x^2 + y^2 + z^2 = 1,$ where ($x, y, z$) are coordinates in~$\mathbb{E}^\pm$~\cite{Pe84}. Despite this flat spacetime approach, the generalization for curved spacetimes shall be direct by means of the tangent bundle.   

Returning to our main exposition, more generally, the direction of any null vector ${\bf v}\in\mathbb{R}^{1,3}$, unless such vector is an element of the  plane defined by the equation $x^0 = 0$, can be represented by two points. Such description results from the intersection of ${\bf v}$ and the hyperplanes  $x^0 = \pm 1$. The future-pointed  ${\bf v}$ is thus  represented by  ($x^1/\|x^0\|, x^2/\|x^0\|, x^3/\|x^0\|$). The inner points of $\mathbb{E}^+$ ($\mathbb{E}^-$) represent the set of future-pointed (past-pointed) light-like directions. 

By considering $\mathbb{E}^+$, by performing a stereographic projection on the Argand-Gauss plane, we obtain a  representation of the union between the set of complex numbers and the point at the infinity, that corresponds to the north pole of  $\mathbb{E}^+$. By defining the complex number \begin{equation} \label{be1}\beta  = \frac{x + iy}{1-z}\,,\end{equation} 
\noindent it yields $\beta \overline\beta  = \frac{x^2 + y^2}{(1-z)^2}$, and consequently \begin{equation} \label{xyz}
x = \frac{\beta  + \overline\beta }{\beta \overline\beta  + 1},\quad y = \frac{\beta  - \overline\beta }{i(\beta \overline\beta  + 1)},\quad z=\frac{\beta \overline\beta  - 1}{\beta \overline\beta  + 1}.
\end{equation} 
\noindent
The correspondence between points of  $\mathbb{E}^+$ and the Argand-Gauss plane is injective if the point $\beta  \sim \infty$ is added to the complex plane, making it to correspond to the north pole with components $(1,0,0,1)$. However, to avoid this point, it is more convenient to  associate a point of $\mathbb{E}^+$ not to a complex number $\beta $, but to a pair of complex numbers \footnote{With the condition that both numbers are not simultaneously zero.} $(\xi, \eta)$, where
\begin{equation} \label{be2}\beta  = \xi/\eta.\end{equation} 
The pairs $(\xi, \eta)$ and $(\lambda\xi, \lambda\eta), \lambda \in \CC$, represent the same point in $\mathbb{E}^+$. Such  components  are called {\it projective coordinates}. These projective coordinates resulting from the stereographic projection of the Riemann sphere on the complex plane, collected in a pair, gives rise to what we call by spinor. It is important to remark that the used Riemann sphere is a quite special one: it results from the intersection of a time constant plane with the (future oriented here) light cone. 

The point $\beta  = \xi/\eta \sim \infty$ corresponds to the point of coordinates ${\xi\choose\eta} = {1\choose 0}$. Eqs. (\ref{xyz})  can be expressed as
\begin{eqnarray} 
\label{xyw}
x = \frac{\xi\overline\eta + \eta\overline\xi}{\xi\overline\xi + \eta\overline\eta},\quad y = \frac{\xi\overline\eta - \eta\overline\xi}{i(\xi\overline\xi + \eta\overline\eta)},\quad z = \frac{\xi\overline\xi - \eta\overline\eta}{\xi\overline\xi + \eta\overline\eta},
\end{eqnarray} explaining the claim of spinors pre-geometric quantities, or the ``square-root of a point''.  
The point $P=(1,x,y,z)$ is an arbitrary point of the light-cone  transversal section with constant time and represents a null  future-pointed direction, that can be represented by any point of the line ${\mathcal{O}} P$. In particular, if a point $R$ is taken in the line  ${\mathcal{O}} P$ by multiplying $P$ by the factor $(\xi\overline\xi + \eta\overline\eta)/\sqrt{2}$, then  $R$ has coordinates 
\begin{eqnarray} \label{xt}x^1 &=& \frac{1}{\sqrt{2}}(\xi\overline\eta + \eta\overline\xi), \quad x^2 = \frac{1}{i\sqrt{2}}(\xi\overline\eta - \eta\overline\xi),\nonumber\\
x^3 &=& \frac{1}{\sqrt{2}}(\xi\overline\xi - \eta\overline\eta), \quad x^0 = \frac{1}{\sqrt{2}}(\xi\overline\xi + \eta\overline\eta).
\end{eqnarray} 
\noindent Notice that contrary to the point  $P$, the point $R$ is not invariant under  $(\xi, \eta) \mapsto (r\xi, r\eta), r \in {\mathbb{R}}$, although it is independent of phases $(\xi, \eta) \mapsto (e^{i\theta}\xi, e^{i\theta}\eta), \theta \in {\mathbb{R}}$. 

Consider now the following complex linear transformation
\begin{equation} 
\begin{array}{l}
\label{tlc}
\xi \mapsto \tilde\xi = \alpha\xi + \mu\eta,\\
\eta \mapsto \tilde\eta = \gamma\xi + \delta\eta,
\end{array}\end{equation} 
\noindent where  $\alpha, \mu, \gamma, \delta \in \CC$ satisfy $\alpha\delta - \mu\gamma \neq 0$, in order to such transformation being invertible. It can be rewritten as 
\begin{equation} \label{mobius}
\beta \mapsto f(\beta ) = \frac{\alpha\beta  + \mu}{\gamma\beta  + \delta},
\end{equation} 
\noindent and named a M\"obius transformation,\index{M\"obius transformation} from the set $\CC\backslash\{-\delta/\gamma\}$ to $\CC\backslash \{\alpha/\gamma\}$. Moreover, if $f(-{\delta/\gamma}) \sim \infty$ and $f(\infty) \sim {\alpha/\gamma}$, then $f$ is an injective function from the complex plane, compactified by the point at the infinity, denoted by  ($\CC \cup \{\infty\}$). 

Hence the space of light-like vectors on Minkowski spacetime is naturally a Riemann sphere. The restricted Lorentz group  ${\mt L}^+$ is, on the other hand, the automorphism group of the Riemann sphere. Eq. (\ref{tlc}) with the condition 
$$\alpha\delta - \mu \gamma = 1$$ 
are called spinor transformations,  \index{spinor transformation}\index{transformation!spinor} where $\beta  = $ $ \xi/\eta $ is related to the null vectors by Eqs. (\ref{xt}), implying that 
\begin{equation} 
\beta  = \frac{x^1 + ix^2}{x^0-x^3} = \frac{x^0-x^3}{x^1-ix^2}.
\end{equation} 
The spinor matrix\index{spinor matrix}\index{matrix!spinor} ${\bf{A}} \in {\rm  SL}(2, \CC)$ is defined as
\begin{equation} \label{mspin}
{\bf{A}} = \left(\begin{array}{cc}
\alpha & \mu\\
\gamma & \delta
\end{array}
\right),\quad {\rm det}\;{\bf{A}} = 1.
\end{equation} 
\noindent Eqs. (\ref{tlc}), with respect to $\bf{A}$, read
$$ \left(
	\begin{array}{c}
	\tilde\xi\\ \tilde\eta
	\end{array}
	\right)
= 
\bf{A} \left(
	\begin{array}{c}
	\xi\\ \eta
	\end{array}
	\right)\,.
$$
The spinor matrices $\{\pm\bf{A}\}$ induce the same transformation of $\beta  = \xi/\eta$.
Eq.(\ref{xt}) yields 
\begin{equation} \label{vesp}\frac{1}{\sqrt{2}}\left(
 	\begin{array}{cc}
 	x^0 + x^3 & x^1 + ix^2\\
 	x^1 - ix^2 & x^0 - x^3
 	\end{array}
 	\right) = 
\left(
 	\begin{array}{cc}
 	\xi\overline\xi\ & \xi\overline\eta\\
 	\eta\overline\xi & \eta\overline\eta
 	\end{array}
 	\right) = 
 \left(
\begin{array}{c}
\xi\\ \eta
\end{array}
\right) (\overline\xi \quad\overline\eta).
\end{equation} 
Hence, up to a factor  $1/\sqrt{2}$, it follows that 
\begin{eqnarray} \left(
 	\begin{array}{cc}
 	x^0 \!+\! x^3 & x^1 \!+\! ix^2\\
 	x^1 \!-\! ix^2 & x^0 \!-\! x^3
 	\end{array}
 	\right) &\mapsto& \left(
 	\begin{array}{cc}
 	\tilde{x^0} \!+\! \tilde{x^3} & \tilde{x^1}\! +\! i\tilde{x^2}\\
 	\tilde{x^1} \!-\! i\tilde{x^2} & \tilde{x^0} \!-\! \tilde{x^3}
 	\end{array}
 	\right)\nonumber\\ &=& \bf{A}\left(
 	\begin{array}{cc}
 	x^0 \!+\! x^3 & x^1 \!+\! ix^2\\
 	x^1 \!-\! ix^2 & x^0 \!-\! x^3
 	\end{array}
 	\right)\bf{A^\dagger}.
 	\end{eqnarray}  
 	\noindent Such transformation acting on the point ${\bf v} = (x^0,x^1,x^2,x^3)$ is real and preserves the light cone structure $(x^0)^2 - (x^1)^2 - (x^2)^2 - (x^3)^2 = 0$. Thus the above relation defines a restricted Lorentz transformation. In fact, the group SL(2,$\CC$) is the two-fold covering of the restricted Lorentz group SO$_+(1,3) \simeq \mathcal{L}^+$.

As mentioned before, all the previous formalization can be performed in a curved space tangent bundle. Some adaptation shall be expected, however, if the base manifold is endowed with a nontrivial topology, for instance, engendered by a black-hole. Indeed, there is no meaning in associate geometrical points to the interior of a given black-hole, evincing thus its existence of nontrivial topology. Nevertheless this means, by its turn, that the tangent bundle itself is not simply connected. Hence, the appearance of different spinors, resulting from different patches in the tangent bundle, are in order. Moreover the spinor dynamics, as partially dictated by the connection, shall also be affected.   

The net effect of the non-trivial topology is labelled by an integer number reflecting, as explained in the previous subsection, the non-triviality of the first cohomology group. This label is recovered as a macroscopic effect by an integration upon a closed curve. From these considerations, the term appearing in the new derivative may be recast into the form 
\begin{eqnarray}
\gamma^\mu\tilde{\nabla}_\mu=\gamma^\mu\nabla_\mu+\frac{1}{2\pi i }\xi^{-1}d\xi,\label{ulti}
\end{eqnarray} where spin and geometric connections are embedded into $\nabla$. Since it must be fulfilled the following requirement   
\begin{eqnarray}
\frac{1}{2\pi i}\oint \xi^{-1}d\xi \in \mathbb{Z}, \label{ma}
\end{eqnarray} we have $\xi=e^{in\theta(x)} \in U(1)$, with $n\in \mathbb{Z}$, in agreement with last subsection. Finally, upon a simple rescaling of the field $\theta(x)$ and bearing in mind that $\gamma^\mu$ is an adequate basis for the orthonormal frame we have 
\begin{equation}
d\theta=\gamma^\mu\partial_\mu \theta,\label{qua}
\end{equation} and the dynamic equation reads 
\begin{eqnarray}
(i\gamma^{\mu}\nabla_\mu+i\gamma^\mu\partial_\mu\theta-m)\tilde{\psi}=0.\label{se}
\end{eqnarray} In the next section we shall pursue an interesting effect connecting the non triviality of the spacetime topology and black holes, in adequate conditions, regarding the Hawking radiation emission rate.

\section{Hawking radiation vs exoticity}

Hawking radiation is one of the main achievements of semi-classical gravity. In fact, it is a widely accepted feature of gravity beyond the Einstein's theory, pointing towards a theory encompassing gravity and quantum mechanics. An interesting and not very explored aspect of the Hawking's result regards its relationship with the exotic structure introduced in the previous section.  As we pointed out, the very existence of a black hole makes the topology of the space-time nontrivial, implying the possible existence of exotic spinors. As mentioned in \cite{uni}, the exotic term does not change the Hawking temperature, however it affects the emission rate from where the temperature is derived. This fact, which is not as spread as it should be, can change the expected life time of black holes. In this section we are going to investigate the possible extremes of the emission rate $\Gamma$ and how it is affected by the exotic term $\theta$. For this reason we must consider the emission of spinors, as their dynamics is affected by the presence of $\theta$. Due to its generalness and compatibility with the Einstein's theory, we have chosen to adopt the emission rate of the Kerr-Newman solution.   

The Hawking radiation emission rate, also interpreted as the tunneling probability \cite{Parikh:1999mf, Angheben:2005rm,Arzano:2005rs, Jiang:2005ba}, is quite straightforwardly derived by using the Hamilton-Jacobi approach to the tunneling method. Such method is based upon the particle description of Hawking radiation,  under the assumption that the emitted particle action does satisfy the relativistic Hamilton-Jacobi equation. It relies on allowing particles to travel along classically forbidden trajectories, from just behind the
horizon onward to infinity. A comprehensive review of the tunneling method can be found in \cite{Vanzo:2011wq}. 

The  Kerr-Newman solution of the Einstein's field equations, in Boyer-Lindquist coordinates $(
t,r,\vartheta,\phi)$, reads
\begin{eqnarray}\nonumber
ds^2=&-\left(\frac{\Delta-a^2\sin^2\vartheta}{\Sigma}\right)dt^2-\frac{2(r^2+a^2-\Delta)a\sin^2\vartheta}{\Sigma}dtd\phi+\frac{\Sigma}{\Delta}dr^2+\Sigma d\vartheta^2+\\ &+\frac{(r^2+a^2)^2-a^2\Delta\sin^2\vartheta}{\Sigma}\sin^2\vartheta d\phi^2,
\end{eqnarray}
where $a=\frac{J}{M}$, $\Sigma=r^2+a^2\cos^2\theta$ and $\Delta=r^2+a^2+{Q^2}-{2Mr}$. We are going to depart from its emission rate of fermionic particles \cite{Vanzo:2011wq, uni}, taking into account the exotic term

\begin{equation}
\Gamma=\exp\left[-4\pi\left(\frac{r_+^2+a^2}{r_+-r_-}\right)(\omega-j\Omega-q\Phi+\dot{\theta})\right].
\end{equation}
Here $r_+$ $[r_-]$ denotes the outer [inner] horizon, given by 
\begin{equation}
r_\pm=M\pm\sqrt{M^2-Q^2-a^2}.
\end{equation}

A preliminary analysis could start from assuming that the horizon radii, as well as the emitted particle and black hole parameters, do not change appreciably in time. In fact, bearing in mind astrophysical black holes, this fact seems quite reasonable. In such a simplified case, however, the condition for the existence of extremes of $\Gamma$ is simply $\ddot{\theta}=0$, that is, the exotic term should be linear. The situation becomes more interesting by relaxing the assumption of constant radii. From now we are going to distinguish two cases, characterized by the existence or not of an external electromagnetic potential $A_\mu$. The tunneling/emission rate will be denoted by $\Gamma_{DN}$ for a vanishing external potential and $\Gamma_{DE}$ for a non vanishing external potential. For Dirac exotic spinors, that is, solutions to the Dirac equation with the additional exotic term, and with no external interaction, the tunneling probability is given, as earlier, by \cite{Vanzo:2011wq, uni}

\begin{equation}\label{gammaDN}
 \Gamma_{DN} = \exp{\left[ -4\pi \left( \frac{r_{+}^2  + a^2}{r_{+} - r_{-}}\right) (\omega - j\Omega - q \Phi + \dot{\theta}) \right]},
\end{equation}
and its first derivative by

\begin{align}\label{derivada1gammaDN}
 \dot{\Gamma}_{DN} = \frac{4 \pi \Gamma_{DN}}{(r_- - r_+)^2} & \left\{ (\omega - j\Omega - q \Phi + \dot{\theta}) \left[(a^2 + r_+^2) (-\dot{r}_- + \dot{r}_+)+\right. \right. \nonumber \\& \left. \left. + 2 (r_-  - r_+) (a \dot{a} + r_+ \dot{r}_+) \right]  + (r_- - r_+) (a^2 + r_+^2) \ddot{\theta} \right\}.
\end{align}
Here we are taking $\omega$, $j\Omega$ and $q\Phi$ as constant in time and  setting $\dot{r}^2, \dot{\theta}^2, \ddot{r} \rightarrow 0$. The later meaning adiabatic variation, i. e., despite the possible extreme emission rate, the radii varies slowly.% In these approximate terms the Einstein equations solutions, ultimately, shall be the same.  

The extreme condition $\dot{\Gamma}_{DN} = 0$ leads to the equation
% 
%\begin{equation}
% \ddot{\theta} + (\omega - j\Omega - q \Phi + \dot{\theta}) \left[ \frac{(\dot{r}_+ - \dot{r}_-)}{(r_- - r_+)} + 2\frac{(a \dot{a} + r_+ \dot{r}_+)}{(r_+^2 + a^2)} \right] = 0,
%\end{equation}
%which can be easily rewritten as
%
\begin{equation}\label{asterisco}
 \ddot{\theta} + (\omega - j\Omega - q \Phi + \dot{\theta}) \frac{d}{dt} \left[ \ln{ \left( \frac{r_+^2 + a^2}{r_+ - r_-} \right) } \right] = 0.
\end{equation}
Note that, in spite of being a particular case, it gives us a second order differential equation for the unknown exotic term $\theta$. Furthermore, we emphasize that this case describes the emission rate of exotic dark spinors (Elko) \cite{uni, Cavalcanti:2015nna,exo}, a prime candidate to describe dark matter, as it interacts only with gravity and the Higgs field  (see \cite{Ahluwalia:2016rwl} and references therein for a general discussion on Elko dark spinors). 

 For the case of exotic Dirac spinors interacting with an $A_\mu$ field, with $\slashed A \equiv \gamma_\mu A^\mu$, one can express the tunnelling probability $\Gamma_{DE}$ for the interacting exotic Dirac spinor as
\begin{equation}\label{gammaDE}
 \Gamma_{DE} = \exp{\left[ -4\pi \left( \frac{r_{+}^2  + a^2}{r_{+} - r_{-}}\right) (\omega - j\Omega - q \Phi + \dot{\theta} + \slashed A) \right]}.
\end{equation}
We point out that, apart from its very nature, the exotic term on the equation above can not be incorporated by the gauge field. As shown in the Appendix \ref{append},  the exotic term induces modifications on the dispersion relation incompatible with the ones due to the gauge field.

From Eqs. (\ref{gammaDN}) and (\ref{gammaDE}) we can write
\begin{equation}
 \Gamma_{DE} = \Gamma_{DN} \mathcal{R},
\end{equation}
where $ \mathcal{R} \equiv \exp{\left[ -4\pi \left( \frac{r_{+}^2  + a^2}{r_{+} - r_{-}} \right) \slashed A \right]}$, following

%\begin{equation}
% \mathcal{R} \equiv \exp{\left[ -4\pi \left( \frac{r_{+}^2  + a^2}{r_{+} - r_{-}} \right) \slashed A \right]}.
%\end{equation}
%Calculating the first time-derivative of $\Gamma_{DE}$, one can find that

\begin{equation}\label{derivada1gammaDE}
 \dot{\Gamma}_{DE} = \dot{\Gamma}_{DN} \mathcal{R} + \Gamma_{DN} \dot{\mathcal{R}}.
\end{equation}
Here the extreme condition gives two possibilities: $\dot{\Gamma}_{DN} = 0$ or $\dot{\Gamma}_{DN} \neq 0$. The first one results in
%Now, doing $\dot{\Gamma}_{DE} = 0$, one has two options: $\dot{\Gamma}_{DN} = 0$ or $\dot{\Gamma}_{DN} \neq 0$. 
%Firstly, we deal with the situation that $\dot{\Gamma}_{DN} = 0$. In this case, if one imposes $\dot{\Gamma}_{DE} = 0$, then it is necessary to have $\dot{\mathcal{R}} = 0$, which leads to
\begin{equation}\label{derivadaRzero}
 \dot{\slashed A} + \slashed A \frac{d}{dt} \left[ \ln{ \left( \frac{r_+^2 + a^2}{r_+ - r_-} \right) } \right] = 0.
\end{equation}
 Comparing Equations (\ref{asterisco}) and (\ref{derivadaRzero}) one finds %it is possible to write down
%
%\begin{equation}
% \frac{\dot{\slashed A}}{\slashed A} = \frac{\ddot{\theta}}{\omega - j\Omega - q \Phi + \dot{\theta}},
%\end{equation}
%which leads to
%
\begin{equation}\label{condicaoDNzero}
 \slashed A  - \dot{\theta} = \omega - j\Omega - q \Phi.
\end{equation}
On the other hand, the condition $\dot{\Gamma}_{DN} \neq 0$ leads to
\begin{equation}\label{condicaoDNnaozero}
 - \slashed A  - \dot{\theta} = \omega - j\Omega - q \Phi.
\end{equation}
%\textcolor{red}{In the process of obtaining Equations (\ref{condicaoDNzero}) and (\ref{condicaoDNnaozero}), we are setting $\slashed A(0) = 0$, i.e., the external field has zero intensity in the initial time $t=0$.}
Again the extreme condition fixates a differential equation for the exotic term, allowing the time variation of $\theta$ being completely determined by the external field.

% We can note a change of behaviour of the exotic term $\dot{\theta}$ in time related to the behaviour of $\slashed A$, in function of having or not having extreme values for the emission of Dirac exotic spinors that are not interacting with any external field. Indeed, note that the right-handed side of Equations (\ref{condicaoDNzero}) and (\ref{condicaoDNnaozero}) are both constants in time. More accurately, let us call $DN-$particles the ones related to the Dirac exotic spinor fields that has no interaction with $\slashed A$, and suppose the situation that $\slashed A$ is increasing with time. Then, if it happens at the moment of extreme emission of $DN-$particles, the exotic term $\dot{\theta}$ also increases, although, on the other hand, if the emission of $DN-$particles are not in its extreme situation, then $\dot{\theta}$ decreases. This relates the dynamics of the external field $\slashed A$ to the behaviour of the exoticiness of the spacetime itself, in the case of this particular Kerr-Newman black hole situation.
 
 Our next step is to investigate the second derivatives of the tunnelling rates. %One has no difficult in obtain, by Equation (\ref{derivada1gammaDE}), the following:
% 
% \begin{equation}\label{derivada2gammaDE}
%  \ddot{\Gamma}_{DE} = \ddot{\Gamma}_{DN} \mathcal{R} + 2 \dot{\Gamma}_{DN} \dot{\mathcal{R}} + \Gamma_{DN} \ddot{\mathcal{R}}.
% \end{equation}
 From Equation (\ref{derivada1gammaDN}) follows

 \begin{align}
   \ddot{\Gamma}_{DN} = \frac{4\pi \Gamma_{DN}}{(r_- - r_+)^4} & \left\{ \left[ 4\pi [\mathcal{X} (\alpha + 2\beta) + \sigma \ddot{\theta}] - 2(\dot{r}_- - \dot{r}_+)(r_- - r_+) \right] \right. \nonumber \\ & \times \left. \left[ \mathcal{X}(\alpha + 2\beta)  + \sigma \ddot{\theta} \right] +
   + (r_- - r_+)^2 \left[ 2\ddot{\theta} (\alpha + 2\beta) + \sigma \dot{\ddot{\theta}} \right] \right\},
 \end{align}
with 
\begin{align}
\alpha &\equiv (a^2 + r_+^2)(-\dot{r}_- + \dot{r}_+)\\
\beta &\equiv (r_- - r_+)(a\dot{a} + r_+ \dot{r}_+)\\
\sigma &\equiv (r_- - r_+)(a^2 + r_+^2)\\
\mathcal{X} &\equiv (\omega - j\Omega - q \Phi + \dot{\theta}).
\end{align}
% Suppose $\dot{\Gamma}_{DN} \neq 0$ in the calculations of $\dot{\Gamma}_{DE} = 0$. Then, by Equation (\ref{condicaoDNnaozero}), one has $\mathcal{X} + \slashed A = 0$, which straightforwadly leads, by Equation (\ref{gammaDE}), to $\Gamma_{DE} = 1$, i.e., in this case we have maximum emission of Hawking radiation for $DE-$particles, with $100\%$ of tunnelling probability.
Starting from the case $\dot{\Gamma}_{DN} = 0 = \dot{\Gamma}_{DE}$ we define
 \begin{align}
  \ddot{\tilde{\Gamma}}_{DE} \equiv & \ddot{\Gamma}_{DE} \Big\rvert_{\dot{\Gamma}_{DN}, \dot{\Gamma}_{DE} = 0},\\
  \ddot{\tilde{\Gamma}}_{DN} \equiv & \ddot{\Gamma}_{DN} \Big\rvert_{\dot{\Gamma}_{DN}, \dot{\Gamma}_{DE} = 0},
 \end{align}
and after some simple yet lengthy calculations, noticing that $\dot{\ddot{\theta}} = \ddot{\mathcal{X}}$ and $\ddot{\theta} = \dot{\mathcal{X}}$, on finds
 
%  \begin{align}\label{derivada2gammaDEDNdoisextremos}
%   \tilde{\ddot{\Gamma}}_{DE} = & \frac{4\pi}{(r_- - r_+)^2} \left[ \mathcal{F}(\Gamma_{DN},\mathcal{X},\slashed A) + \mathcal{F}(\mathcal{R},\slashed A,\mathcal{X}) \right],\\
%   \tilde{\ddot{\Gamma}}_{DN} = & \frac{4\pi}{(r_- - r_+)^2} \mathcal{F}(\Gamma_{DN},\mathcal{X},0),
%  \end{align}

  \begin{align}\label{derivada2gammaDEDNdoisextremos}
   \ddot{\tilde{\Gamma}}_{DE} = & \frac{4\pi \Gamma_{DE}}{(r_- - r_+)^2} \left[ f(\mathcal{X}) + f(\slashed A) \right],\\
   \ddot{\tilde{\Gamma}}_{DN} = & \frac{4\pi \Gamma_{DN}}{(r_- - r_+)^2} f(\mathcal{X}),
  \end{align}
where  $f(y) \equiv 2\dot{y} (\alpha + 2\beta) + \sigma \ddot{y}.$
%
%\begin{equation}
% f(y) \equiv 2\dot{y} (\alpha + 2\beta) + \sigma \ddot{y}.
%\end{equation}
The signs of $\ddot{\tilde{\Gamma}}_{DE}$ and $\ddot{\tilde{\Gamma}}_{DN}$ are then strongly related to the signs of $f(\mathcal{X})$ and $f(\slashed A)$. 
Note that in this case, by Eq. (\ref{condicaoDNzero}), one has $\mathcal{X} = \slashed A$. Therefore, the sign of $f(\mathcal{X}) = f(\slashed A)$ defines identical signs for $\ddot{\tilde{\Gamma}}_{DE}$ and $\ddot{\tilde{\Gamma}}_{DN}$. This means that, % in the case of $\dot{\Gamma}_{DN} = 0 = \dot{\Gamma}_{DE}$, $DN-$particles and $DE-$particles have coincident extreme radiations, i.e., maximum for both or minimum for both. In other words, 
for our particular Kerr-Newman black hole situation, the presence of an external field $A_\mu$ does not change the general behaviour of the emission extremes of Hawking radiation for an exotic particle, remaining it maximum [minimum] if it was already maximum [minimum]. In order to establish rather the emission rate is maximum or minimum for the  $\dot{\Gamma}_{DN} = 0 = \dot{\Gamma}_{DE}$ case
%
% One sees that, since with $\dot{\Gamma}_{DN} \neq 0$ and $\dot{\Gamma}_{DE} = 0$ we have necessarly $DE-$particles with maximum radiation, the only possibility to reach a minimum tunnelling probability for $DE-$particles is with $DN-$particles in the same situation (i.e., also with minimum radiation): this scenario is $\dot{\Gamma}_{DN} = 0 = \dot{\Gamma}_{DE}$, and now we will go further on this, checking the conditions for maximum and minimum for this case by analysing the possible signs of $f(\mathcal{X}) = f(\slashed A)$.
% %
% Firstly, 
 we must know the signs of the $f(y)$ coefficients. %One has no difficult on verifying that $\sigma < 0$. On the other hand, 
 Even though  $\alpha < 0$ and $\beta > 0$, there is no general sign for $\alpha + 2\beta$. In fact, with $z \equiv M^2 - (\frac{J}{M})^2 - q^2 > 0$ and $\frac{dM}{dt} < 0$, we can write
 
 \begin{align}\label{termoescroto}
  \alpha + 2\beta = \frac{dM}{dt} & \left[ \left( \frac{4M^2}{\sqrt{z}} + 4 - \frac{2q^2}{\sqrt{z}} - 8M\sqrt{z} \right) \left( M + \frac{J^2}{M^3} \right) + \right. \nonumber \\
  & + \left. \left( - 8M^2\sqrt{z} - 12Mz - 8z^{\frac{3}{2}} - 8M^3 + \frac{4J^2 \sqrt{z}}{M^3} - \frac{8J^2}{M} \right) \right].
 \end{align}
Since we do not have constraints on $J$, $M$ and $q$ other then $z > 0$, the sign of (\ref{termoescroto}) is undetermined. However, we can set the conditions that would lead to maximum or minimum values of the tunnelling probabilities, in terms of the signs of the derivatives of $\dot{\theta}$ and $\slashed A$. We have summarized all the possible cases below (adopting the notation $y \equiv \mathcal{X} = \slashed A$):
 \begin{enumerate}

 \item  $\alpha + 2\beta < 0$:
  \begin{enumerate}
  \item $\dot{y} < 0$, $\ddot{y} < 0$: $f(y) > 0$, minimum emission rate;
  \item $\dot{y} > 0$, $\ddot{y} > 0$: $f(y) < 0$, maximum emission rate;
  \item $\dot{y} > 0$, $\ddot{y} < 0$: both cases are possible;
  \item $\dot{y} < 0$, $\ddot{y} > 0$: both cases are possible.
 \end{enumerate}
 
 \item $\alpha + 2\beta > 0$:
  \begin{enumerate}
  \item $\dot{y} < 0$, $\ddot{y} < 0$: both cases are possible;
  \item $\dot{y} > 0$, $\ddot{y} > 0$: both cases are possible;
  \item $\dot{y} > 0$, $\ddot{y} < 0$: $f(y) > 0$, minimum emission rate;
  \item $\dot{y} < 0$, $\ddot{y} > 0$: $f(y) < 0$, maximum emission rate.
 \end{enumerate}

 \end{enumerate}
Finally, there are two conditions ruling these scenarios: all maximum emission rate are related to the condition $\displaystyle \frac{\dot{\sigma}}{\sigma} < - \frac{1}{2} \frac{\ddot{y}}{\dot{y}}$, while the minimum emission rates are linked to $\displaystyle \frac{\dot{\sigma}}{\sigma} > - \frac{1}{2} \frac{\ddot{y}}{\dot{y}}$. Note also that $\dot{\sigma} = \alpha + 2\beta$. The same results concerning the behaviour of the tunnelling probabilities are valid for all asymptotically flat black hole solution in general relativity,  since setting $q = 0$ and/or $J = 0$ does not affect any of our conclusions under the same assumptions ($\omega$, $j\Omega$ and $q\Phi$ constants in time and $\dot{r}^2, \dot{\theta}^2, \ddot{r} \rightarrow 0$). Also, note that in all calculations we did not fixed any restriction to $\slashed A$ rather than $\slashed A(0) = 0$.

\section{Conclusion}

In this paper we investigated the relationship between exotic spinorial structures and the emission rate of spinors by asymptotically flat black holes from general relativity. We established general conditions for the emission rate being extreme in two different cases, considering the existence or not of an external field $A_\mu$. From general conditions for the emission rate extremes we found differential equations for the exotic term. %We pointed out that the absence of external field is particularly interesting due to the fact that this case describes dark spinors (Elko), as it does not interact with external fields other than gravity and Higgs.

 In particular, we found the dynamics of the exotic term $\dot{\theta}$ to the dynamics of $A_\mu$, depending on having or not extreme values for the emission of Dirac exotic spinors. Indeed, note that the right-handed side of Eqs. (\ref{condicaoDNzero}) and (\ref{condicaoDNnaozero}) are both constants in time. More accurately, let us call $DN-$particles the ones related to the Dirac exotic spinor fields that have no interaction with $A_\mu$, and suppose the situation that $\slashed A$ is increasing with time. Then, if it happens at the moment of extreme emission of $DN-$particles, the exotic term $\dot{\theta}$ also increases. On the other hand, if the emission of $DN-$particles is not in its extreme state, the exotic term $\dot{\theta}$ decreases. This relates the dynamics of the external field $A_\mu$ to the exoticiness of the spacetime itself, in the case of any black hole solution in general relativity.

 Moreover, in the case of $\dot{\Gamma}_{DN} \neq 0$ and $\dot{\Gamma}_{DE} = 0$ simultaneously, the Eq. (\ref{gammaDE}) results in $\Gamma_{DE} = 1$, i.e., in this case we have maximum emission of Hawking radiation for $DE-$particles, with $100\%$ of tunnelling probability. % In other words, having the extreme case for $DE-$particles without the conditions for extreme emission rate of $DN-$particles leads necessarily to a scenario with $DE-$particles with maximum radiation.
  Therefore, the only possibility to reach a minimum tunnelling probability for $DE-$particles is with $DN-$particles in the same situation (i.e., also with minimum radiation): this scenario happens if $\dot{\Gamma}_{DN} = 0 = \dot{\Gamma}_{DE}$. In order to check the conditions for maximum and minimum for this case by analysing the possible signs of $f(\mathcal{X}) = f(\slashed A)$, we have found that the sign of (\ref{termoescroto}) is undetermined. %, i.e., at first both scenarios are possible (maximum and minimum tunneling probabilities).  
  Although, we have found a condition to separate these situations in terms of inequalities: maximum emission rate is happening when $\displaystyle \frac{\dot{\sigma}}{\sigma} < - \frac{1}{2} \frac{\ddot{y}}{\dot{y}}$, while minimum emission rate is related to the inequality with opposite sign. The conditions that lead to these relations are displayed in the (a-d) cases for the possible signs of $\alpha + 2\beta$. In this sense, the maximum and minimum emission rates are constrained by the exotic term. %(which is, in this case, the same of the external electromagnetic potential) 
%  encoded by the signs of their first and second derivatives.
 
 Then, by means of Hawking radiation emission rates, we have found relations between the black hole's parameters mass, charge and angular momentum (via the $\dot{\sigma} = \alpha+2\beta$) with the spacetime topology (encoded in the exotic term $\dot{\theta}$) and the behaviour of external fields near to the black hole itself. It can help us to give a step ahead into the understanding of the yet not well known (up to our knowledge) behaviour of the exotic topological term, potentially enlightening the full dynamics of exotic spinors.

% ------------------------------------------------------------------------

\appendix

\section{Exotic dispersion relations}\label{append}

There is an important history in similarities and differences concerning additional vectorial terms in Dirac equation. In this Appendix we would like to evince one more point regarding this history. In fact, we shall start from the old, but important, fact that every connected group have an unitary representation. This simple observation is crucial to understand the root of the difference between the exotic term appearing in the Dirac equation and the gauge interacting field. 

As well known, the gauge group of the electromagnetism is $U(1)$ and, hence, its elements are disposed as $e^{i\Lambda(x)}$ with $\Lambda \in \mathbb{R}$. Now, departing from the usual Dirac equation $(i\gamma^\mu \partial_\mu - m)\Psi = 0,$
and performing the transformation $\partial_\mu \rightarrow \partial_\mu - iA_\mu$, one finds $i\gamma^\mu \partial_\mu \rightarrow i\gamma^\mu \partial_\mu + \gamma^\mu A_\mu$, which turns the Dirac equation, along with the exotic extra-term, into
\begin{equation}\label{exoticdirac}
 (i\gamma^\mu \partial_\mu + \gamma^\mu A_\mu + i\gamma^\mu \partial_\mu \theta - m)\Psi = 0.
\end{equation} Performing a gauge transformation $A_\mu \rightarrow A_\mu - \partial_\mu \Lambda$, it would be necessary to have $\Lambda = i\theta$ to absorb the exotic term into a gauge transformation. However $\theta \in \mathbb{R}$, as it can be read from the very construction of the exotic term, thus it is not possible to have the exotic term acting as a shift on the gauge field.

The exotic term and the gauge field have different nature themselves. In fact, both are vectorial fields, but there is not a gauge tuning which encompass the exotic term.

%--------------------------------------

Besides their very different origins, the gauge field and the exotic term lead to different dispersion relations, emphasizing that $\partial_\mu \theta$ is a vector but not a gauge field. From now on, we shall investigate a simple toy model evincing such a point. 

%
%\vspace{1cm}
%
Using $\gamma^\mu (A_\mu + i\partial_\mu \theta) \equiv \gamma^\mu \tilde{A}_\mu$ in the Eqn. (\ref{exoticdirac}) leads to
\begin{equation}\label{diractilde}
 (i\gamma^\mu - m + \gamma^\mu \tilde{A}_\mu) \Psi = 0.
\end{equation}
Imposing a plane wave solution
\begin{equation}
 \Psi = \omega(p) e^{\mp i px},
\end{equation}
Eqn. (\ref{diractilde}) provides
\begin{equation}\label{diracplane}
 (\pm \slashed p - m + \gamma^\mu \tilde{A}_\mu) \omega(p) = 0.
\end{equation}
Thus, in order to have nontrivial solution, there cannot exists an inverse for $(\pm \slashed p - m + \gamma^\mu \tilde{A}_\mu)$. Certainly, one can write
\begin{equation}
 (\pm \slashed p - m + \gamma^\mu \tilde{A}_\mu)^{-1} = \frac{(\pm \slashed p + m + \epsilon \gamma^\mu \tilde{A}_\mu)}{(\pm \slashed p - m + \gamma^\mu \tilde{A}_\mu)(\pm \slashed p + m + \epsilon \gamma^\nu \tilde{A}_\nu)},
\end{equation}
with $\epsilon$ being a real parameter. Then, rewriting the denominator, we find the condition
\begin{equation}\label{impose}
 p^2 - m^2 \pm 2 \epsilon p^\mu \tilde{A}_\mu + (1-\epsilon)(m\gamma^\mu \tilde{A}_\mu \pm \gamma^\mu \tilde{A}_\mu \slashed p) + \epsilon \tilde{A}^\mu \tilde{A}_\mu = 0
\end{equation}
to be sufficient to get nontrivial solution for Equation (\ref{diracplane}). Now we are able to consider some particular cases of the above equation.

\begin{itemize}
\item $\tilde{A}_\mu \rightarrow 0$:
\end{itemize}

In this case, the imposition (\ref{impose}) leads to $p^2 = m^2$, i.e., the usual dispersion relation $E^2 = \vec{p}^2 + m^2$, expected for plane waves.

\begin{itemize}
\item $\partial_\mu \theta \rightarrow 0$ (or, equivalently, $\theta$ is constant):
\end{itemize}

Here we have $\tilde{A}_\mu \rightarrow A_\mu$. With $\epsilon = 1$, one has $p^2 = m^2 \mp 2p^\mu A_\mu - A^2$, which leads, using the Coulomb gauge, to
\begin{equation}\label{dispersionGauge}
 E^2 = m^2 + (\vec{p} \pm \vec{A})^2,
\end{equation}
revealing an unusual spectrum for the plane wave solution. Either way, it is a feasible observation signature typical for plane wave like fermions interacting with the electromagnetic field, without exotic term.

\begin{itemize}
\item $A_\mu \rightarrow 0$:
\end{itemize}

This case leads to
\begin{equation}
 p^2 - m^2 + \epsilon \partial^\mu \theta \partial_\mu \theta + i [\pm 2\epsilon p^\mu \partial_\mu \theta + (1-\epsilon)\gamma^\mu \partial_\mu \theta (m \pm \slashed p)] = 0,
\end{equation}
and there are no choices for $\epsilon$ providing a real dispersion relation. Notice, however, that with $\epsilon = 1$ we have
\begin{equation}
 p^2 - m^2 + \partial^\mu \theta \partial_\mu \theta \pm 2i p^\mu \partial_\mu \theta = 0
\end{equation}
and a topology that generates a $\theta$ such that\footnote{A sufficient condition for $p^\mu \partial_\mu \theta = 0$ is $\partial_\mu \theta \sim a_\mu$, the 4-acceleration.} $p^\mu \partial_\mu \theta = 0$ can lead to a relation of the type $p^2 = m^2 - \partial^\mu \theta \partial_\mu \theta$. In this situation, we have
\begin{equation}
 E^2 = \vec{p}^2 + m^2 - \partial^\mu \theta \partial_\mu \theta.
\end{equation}

Moreover, due to the constraint $p^\mu \partial_\mu \theta = 0$, we find that the relation $E^2 = \frac{(\vec{p} \boldsymbol{\cdot} \vec{\nabla}\theta)^2}{\dot{\theta}^2}$ must be considered and, eventually, used to constraint the exotic term.

The relevant point here is that the dispersion relation
\begin{equation}
 E^2 = \vec{p}^2 + m^2 - \dot{\theta}^2 +  \vec{\nabla} \theta \boldsymbol{\cdot} \vec{\nabla} \theta
\end{equation}
obtained in this case is different from that obtained in Equation (\ref{dispersionGauge}), which had only took into account the gauge field interaction.

Finally, the case with all terms (gauge field and exotic) may also be analysed, leading to a corresponding and different dispersion relation.
%
%\begin{itemize}
%\item 
%\end{itemize}
%With $A_\mu, \partial_\mu \theta \neq 0$ and $\epsilon = 1$, we have
%\begin{equation}
%p^2 - m^2 \pm 2 p^\mu A_\mu + A^2 + \partial^\mu \theta \partial_\mu \theta + 2i\partial_\mu \theta (A^\mu \pm p^\mu) = 0.
%\end{equation}
%In order to have a real dispersion relation, let us impose $\partial_\mu \theta (A^\mu \pm p^\mu) = 0$ on the above equation, turning it into
%\begin{equation}
 %E^2 = m^2 + (\vec{p} \pm \vec{A})^2 - \partial^\mu \theta \partial_\mu \theta.
%\end{equation}
%This imposition can lead to $ E^2 = \frac{[\vec{\nabla} \theta \boldsymbol{\cdot} (\vec{A} \pm \vec{p})]^2}{\dot{\theta}^2}$.
%
Despite the simplicity, such analysis demonstrates that there exists a possible observable difference between an interacting and a topological Dirac operator.

% ------------------------------------------------------------------------

\subsection*{Acknowledgment}
The authors are grateful to Professor Rold\~ao da Rocha for bringing us together into the spinorial issues. JMHS thanks to CNPq for partial support. RCT thanks the UNESP-Guaratinguet´a Post-Graduation program and CAPES. DB thanks to CAPES for the financial support.

% ------------------------------------------------------------------------
\end{document}